 \documentclass[12pt,preprint]{aastex}

\newcommand{\kms}{km\thinspace s$^{-1}$}

\newcommand{\Msun}{M$_\odot$}
 
\begin{document}

\title{Filaments as Possible Signatures of Magnetic Field Structure in Planetary Nebulae$^1$}


\author{P. J. Huggins \& S. P. Manley}

\affil{Physics Department, New York University, 4 Washington
Place, New York NY 10003}

\email{patrick.huggins@nyu.edu, sm1006@nyu.edu} 
\altaffiltext{1}{Based on observations made with the NASA/ESA Hubble
     Space Telescope, obtained from the data archive at the Space
     Telescope Institute. STScI is operated by the association of
     Universities for Research in Astronomy, Inc. under the NASA
     contract NAS 5-26555.}


\begin{abstract}
We draw attention to the extreme filamentary structures seen in
high-resolution optical images of certain planetary nebulae.  We
determine the physical properties of the filaments in the nebulae
IC~418, NGC~3132, and NGC~6537, and based on their large
length-to-width ratios, longitudinal coherence, and morphology, we
suggest that they may be signatures of the underlying magnetic
field. The fields needed for the coherence of the filaments are
probably consistent with those measured in the precursor circumstellar
envelopes. The filaments suggest that magnetic fields in planetary
nebulae may have a localized and thread-like geometry.

\end{abstract}


\section{Introduction}

High-resolution imaging at optical wavelengths, especially with the
Hubble Space Telescope (HST), has revealed a remarkable degree of
complexity in the morphology of planetary nebulae (PNe).  This
includes both large and small scale structures such as multiple arcs,
bubbles, bicones, point symmetric knots and bullets, tori, ansae, and
globules (see, e.g., the recent volume edited by Meixner et al.\
2004). The multiple arcs are formed in the precursor Asymptotic Giant
Branch (AGB) phase (e.g., Mauron \& Huggins 2000), but most of the
other structures are thought to result from dynamical processes
that take place in the circumstellar gas during the rapid transition
from the AGB to the young PN phase.

The possible role of magnetic fields in forming the complex structures
seen in PNe is an area of ongoing debate. From a theoretical point of
view, magnetic fields close to the central stars of PNe have been
widely discussed in the context of models for launching bipolar jets
which produce point symmetries, either from single stars or accretion
disks in binaries (e.g., Frank \& Blackman 2004; Garcia-Segura, Lopez,
\& Franco 2005). These models, however, are not currently very well
constrained. The role of magnetic fields farther from the central
stars has not been explored in detail. Early ideas were summarized by
Pascoli (1985) who noted a possible connection between filaments and
magnetic fields. More recent developments are described by Soker
(2002; 2005) and references therein.

There are a growing number of observations of magnetic fields in PNe,
but they do not yet form a complete picture.  Magnetic fields in the
central stars of PNe have very recently been detected for the first
time, with kG fields (Jordan et al. 2005). In the extended nebulae,
magnetic fields have been measured in a few young/proto-PNe using the
polarization of maser spots in remnant circumstellar molecular gas
(e.g., Miranda at al. 2001; Bains et al. 2004). More extensive
measurements of maser spots have been made in the precursor
circumstellar envelopes of AGB OH/IR stars (e.g., Vlemmings et al.\
2002; Cotton et al. 2004). The masers extend from close to the stellar
surface (in the case of the SiO masers in AGB stars) out to $\sim
10^{16}$~cm (for the OH masers), and in well observed cases the
strength of the magnetic field is found to be significant, e.g., the
plasma $\beta$ parameter is $\la 1$, where $\beta = 8 \pi P_{\rm
g}/B^2$, $P_{\rm g}$ is the gas pressure, and $B$ is the magnetic
field strength. However, the highly localized nature of the maser
spots leaves uncertain the overall importance and geometry of the
magnetic field. The orientation of the field can, in principle, be
independently measured using the polarization of radiation from
circumstellar dust grains.  Sub-millimeter observations of the
polarization in the young PN NGC~7027 and the proto-PN AFGL~2688 have
been reported by Greaves (2002), although the angular resolution of
the observations (15\arcsec) is too low to resolve the geometry.

In spite of the recent wealth of high-resolution imaging of PNe with
the HST, there have so far been no compelling suggestions that these
images provide direct evidence of magnetic field structures in the
nebulae. Garcia-Segura et al. (2001) have proposed that the multiple
arcs formed in the AGB phase and seen around PNe are caused by changes
in the magnetic pressure, but this is still an open question.  In
contrast, in the case of the Sun, it is well known that high
resolution images of the extended solar atmosphere provide some of the
most direct and powerful diagnostics of the solar magnetic field.
Although magnetic fields may be important close to the central stars
in PNe, the commonly assumed situation is that the fields farther out
are too weak to directly affect the structure of the nebulae on the
size scales observed. For example, for a PN at 500~pc, an offset of
10\arcsec\ from the central star corresponds to a radial distance
$\sim 5000\,R_*$, where $R_*$, the stellar radius, is taken to be
1~A.U.\ for the AGB proto-PN transition. At this distance from the
central star, the magnetic field is extremely small for a dipole
radial variation ($B\sim r^{-3}$) and any plausible stellar surface
field.  However, a dipole variation may not be correct. The
observations of the circumstellar envelopes of OH/IR stars indicate
that the radial variation of the magnetic field in the maser spots may
be less steep, $B\sim r^{-2}$ (Vlemmings et al.\ 2002; Vlemmings at
al. 2005), and, under certain conditions, theoretical considerations
suggest that the fall-off could be shallower, $B\sim r^{-1}$ (e.g.,
Pascoli et al. 1992; Chevalier \& Luo 1994).  Thus, the possible role
of magnetic fields in the extended structure of PNe is an open
question.

To explore this idea, we have examined high-resolution optical images
obtained with the HST in order to search for possible signatures of
the structural effects of magnetic fields in PNe.  In this paper we
present examples where we believe there is substantial evidence that
the observed structure is the result of the underlying magnetic
field. We estimate the magnetic field strengths, and discuss the
implications.

\section{Filaments as Signatures of the Magnetic Field}

HST images of PNe, including the PNe discussed here, reveal a wealth
of small scale structure, which is usually interpreted as the result
of hydrodynamic processes involving the interactions of hot bubbles or
jets with the surrounding circumstellar gas. The question we raise
here is: if a magnetic field is present in the extended nebula, either
advected, or already present in the circumstellar gas, what would be a
morphological signature of its presence?

There is no straightforward answer to this question because there are
no detailed, robust theoretical predictions. There are, however,
general constraints from well observed examples and MHD simulations of
related astrophysical environments.  The dominant structure arising
from the presence of a magnetic field in these cases is filamentary.
Filaments are major structural features of the warm interstellar
medium and cool molecular clouds. Most importantly there is the well
observed case of the Sun where elongated magnetic structures --
magnetic filaments or flux ropes -- are the basic building blocks of
the solar magnetic field. They are formed inside the Sun by the solar
dynamo, they emerge through the solar surface in sunspots, they
provide the complex helical structure of prominences, and form a
component of the solar wind (see, e.g., Russell, Priest, \& Lee 1990).

The magnetic field along a filament provides a connectivity in the gas
that is not otherwise present. For this reason long filaments can
remain coherent when accelerated non-uniformly; they can be twisted
along their length; they can exhibit kinks or macroscopic writhe
forming helices; and they can drape around and tangle with each other,
until they re-connect.  These configurations are probably not easily
produced by purely hydrodynamic processes. Two-dimensional simulations
by Dwarkadas \& Balick (1998) suggest that hydrodynamics can form
types of filamentary structure in PNe, although those produced in the
simulations typically appear short and oriented along the direction of
wind interaction, unlike those discussed in this paper.  We regard
extreme filamentary structures, especially if they exhibit the
characteristics described here, as likely signatures of the underlying
magnetic field.  In the following sections, we present examples of
such filaments in three PNe.

\section{Observations}

The PNe that we consider, NGC~3132, IC~418, and NGC~6537, are 
all well known and have been studied at many wavelengths. Table~1
lists some  basic parameters for orientation and later
discussion. The distances ($d$), angular diameters
($\theta_{\rm D}$), and expansion velocities ($V_{\rm exp}$) are taken
from Acker et al. (1992) or Gathier (1987), and the stellar
temperatures ($T_*$) and electron densities ($n_{\rm e}$) are taken
from Pottasch (2000), Pottasch, Bientema, \& Feibelman (2000),
Pottasch et al.\ (2004), Preite-Martinez et al. (1991), and Gathier
(1987).  The values of $n_{\rm e}$ are characteristic of values found
from emission line ratios and from the intensities of the Balmer lines
or the radio continuum. The expansion timescales are given by $t_{\rm
exp} = \theta_{\rm D} d/ (2 V_{\rm exp})$.

The three PNe are young or middle-aged (2,000--4,000~yr), but the
wide range in temperature of the central stars indicates different
evolutionary stages and/or different core masses. The morphologies are
also different (see Section 4).  All are relatively nearby, so that
the features we describe are likely to be characteristic of a wide
range of more distant objects in the PN population.

The observations that we discuss were obtained with the Wide Field and
Planetary Camera (WFPC2) on board the HST. The data were retrieved
from the HST archives, and have been processed through standard
pipelines. Details of the exposures are given in Table 2. The pixel
size is 0\farcs{05} for the images of IC~418, and 0\farcs{10} for
those of NGC~3132 and NGC~6537.  Images of all three PNe have been
reported in HST press releases, and NGC~3132 and IC~418 appear in the
Hubble Heritage gallery. However, to our knowledge, none of these data
have been described in detail in the astronomical literature. Our
objective is not to present the complete HST data sets, but to focus
on specific features of the nebulae.

\section{Properties of the Filaments}

\subsection{NGC 3132}

NGC~3132 is known as the Southern Ring Nebula because it has a
morphology that is somewhat similar to its northern counterpart
NGC~6720.  An overview of its structure is shown in Fig.~1. The
feature of interest is the bar or ridge which forms a chord across the
main ring, and appears to be a remnant of a thin torus. It lies nearly
vertical, to the left of the bright star near the center in Fig.~1.

High resolution HST images (see Table~2) of the upper part of this
feature in [\ion{O}{3}] $\lambda$5007 and [\ion{N}{2}] $\lambda$6583
are shown Fig.~2. The left hand panels are direct images with
logarithmic intensity scales (shown next to each) to cover the full
dynamic range of the data. The right hand panels are unsharp mask
images that reduce background variations and emphasize the small scale
structure. The unsharp mask versions were formed from the ratio
$I_o$/($I_{\rm s} + 0.05I_{\rm o}$) where $I_{\rm o}$ is the original
image and $I_{\rm s}$ is a smoothed image, made using a Gaussian of
FWHM 14 pixels.

In the [\ion{O}{3}] $\lambda$5007 image, the bar is seen mainly in
absorption by dust grains.  The bar lies in front of most of the
\ion{O}{3} ionization zone along the line of sight, and the
[\ion{O}{3}] $\lambda$5007 emission serves as background illumination
for absorption by the dust in the bar. In contrast, the \ion{N}{2} 
ionization zone extends farther from the central star and includes the
bar region, so that the image in [\ion{N}{2}] shows emission from the
bar, as well as dust absorption.

Inspection of the images shows that the bar is not
monolithic. It consists of several related, thin filaments. One
prominent filament, seen in absorption in the upper left of the
[\ion{O}{3}] image, is relatively straight; it turns through an angle
of $\sim 50\arcdeg$ at the location of the knot which lies just above
the center of the image, and continues straight along the length of
the bar. Parts of the edges of this filament are seen bright in
[\ion{N}{2}].  A second prominent and sinuous filament seen in the
[\ion{O}{3}] image in the upper right, connects with the knot, and
appears to spiral around the taut filament along the length of the
main bar. A third filament, bright in [\ion{N}{2}] and distinctly
kinked, connects with the knot from the top, although the geometry of
its continuation is unclear.  Other, less prominent filaments,
including a distinct loop above the knot, connect to these
main features.

The well-defined absorption in the [\ion{O}{3}] $\lambda$5007 image
makes it very useful for determining the properties of the filaments.
Horizontal intensity cuts across the image are shown in Fig.~3 to
illustrate the widths and depths of the absorption.  The widths (FWHM)
of the more prominent strands are $0\farcs2$ -- $0\farcs6$ and
somewhat larger at the main knots or kinks.  At the
distance of NGC~3132 (Table~1), a representative width of $0\farcs4$
corresponds to $3.6\times 10^{15}$~cm. This subtends an angle of only
$\sim 1\arcdeg$ at the central star. The longest filament can be
traced more than a quarter of the way around the nebula, and has a
length/width ratio of $\ga 60$.

We estimate the density in the filaments from the depths of the dust
absorption.  The peak absorption ranges up to $\sim$50--70\% for the
most prominent kinks and knots, and we adopt a value of 20\% as
representative of a typical, single filament.  At the wavelength of
the [\ion{O}{3}] $\lambda$5007 line, this corresponds to a color
excess $E$(B$-$V) of 0.08, and a hydrogen column density
{$N$(\ion{H}{1} + H$_2)$} of $5\times 10^{20}$~cm$^{-2}$, using the
relation {$N$(\ion{H}{1} + H$_2$) = 5.8 $\times 10^{21} E({\rm B-V})$}
from Bohlin, Savage, \& Drake (1978).  For a line of sight depth equal
to the typical width of a filament given above, the density is $1.3
\times 10^5$~cm$^{-3}$, much higher than the average electron density
given in Table~1. The gas in the filaments can be expected to range
from ionized at the surface to molecular in the core. H$_2$ emission
can be seen along the bar in the data of Allen et al.\ (1997),
although it is not well resolved. The situation is probably like that
in the globules of the Helix Nebula (Huggins et al. 2002). There is
likely to be a range in temperature from the surface to the core, and
if we adopt an intermediate state of neutral atomic gas at a
temperature of 100~K, the pressure is $1.8\times
10^{-9}$~dyne\,cm$^{-2}$.

The properties of the filaments in NGC~3132 described above place
strong constraints on possible formation scenarios. Based on purely
hydrodynamic considerations the matter ejected in the precursor AGB
wind is expected to be smoothed out on the smallest angular sizes
scales in the acceleration region (Huggins \& Mauron 2002). The
filaments lie at larger distances from the central star and they are
oriented tangentially. They cannot be produced at their current
location as wind swept tails, and are unlikely to be produced by
hydrodynamic instabilities such as the Rayleigh-Taylor or nonlinear
thin-shell instabilities as modeled by Dwarkadas \& Balick (1998),
where the structure is typically short and oriented along the
direction of the wind interaction.  Tangential optical filaments are
pervasive in evolved supernova remnants, and are generally interpreted
as thin sheets of emission, seen edge-on (Raymond et al. 1988). In
PNe, similar effects are commonly seen as limb-brightened bubble
surfaces, but other linear features (see, e.g., O'Dell et al. 2002)
may be similar to those discussed here.  In NGC~3132 the images show
that the filaments are long, thin physical structures, and together
they form a vestigial ring or torus. They might arise from some
shearing instability, but the extreme thread-like character of the
filaments together with their local kinked and looped geometry lead us
to propose that their structure is dominated by a magnetic field. The
most likely scenario is that they were formed in the early development
of the PN close to the star, possibly in a magnetized torus in the
proto-PN phase, and their coherence has been maintained by the
influence of the longitudinal field.

\subsection{IC 418}

IC~418 is a well-studied, compact PN with a high surface
brightness in the Balmer lines and the radio continuum. An overview of
the structure in H$\alpha$ is shown in Fig.~4. In addition to the
classic elliptical morphology with enhanced intensity along the minor
axis, it exhibits complex, small scale structure in the form of
filaments and blisters. This is seen more clearly in the unsharp mask
image (made in the same way as described in the previous section)
shown in the top left panel of Fig.~5.

The filaments in IC~418 are brighter than the background nebula
emission in H$\alpha$, which indicates that they consist of higher
density gas. Individual filaments can be traced over distances
comparable to the diameter of the nebula, and in certain areas they
form a semi-regular grid pattern.  The blistered regions
probably occur where the ionized gas in the interior of the nebula has
pushed through the pattern. The globally co-ordinated structure
indicates that the grid pattern was formed by ejection from the center
over time, with a gradual change in the orientation of the ejection.
A more detailed model will be reported in a separate paper.

The thread-like nature of the filaments and their local geometry can
be seen in the close-up unsharp mask images in Fig.~5. The two panels
on the right show the same field in H$\alpha$ and [\ion{N}{2}]
$\lambda$6583, centered below and to the left of the central star (as
seen in the full image) where the pattern is clearest. The bottom-left
panel shows the enlarged center of this field in H$\alpha$. The images
have been smoothed with a Gaussian of width (FWHM) 2 pixels to improve
the visibility without significant loss of detail. The pattern in the
images is mostly formed by enhanced emission from the filaments,
although there are also parts of filaments where absorption or
geometric effects due to the oblique illumination by the central star
give a diminution of the light. The filaments can be expected to be at
different distances along the line of sight, but many span a large
angular extent and reach the edges of the nebula, which indicates that
they are typically near the surface. The bright feature seen in the
top right corner of the intermediate scale H$\alpha$ image is part of
an interior ring around the central star, and is not seen in
[\ion{N}{2}] because the \ion{N}{2} ionization zone begins farther
from the star. Most of the other features can, however, be traced
point by point with slight differences in H$\alpha$ and [\ion{N}{2}],
which argues strongly that they trace real structural elements. The
HST image in [\ion{O}{3}] $\lambda$5007 (not displayed here) shows a
diffuse nebula with the inner ring and a few filaments in emission,
together with evidence of dust absorption, but not sufficiently
clearly for systematic measurements as in NGC~3132.

The typical widths of bright individual strands (measured on the
original H$\alpha$ image) are $\sim 3$ pixels (${0}\farcs{14}$), which
corresponds to $2\times10^{15}$~cm at the distance of IC~418. The
strands subtend an angle of only $\sim 1\arcdeg$ at the central star, and
those that can be traced over more than half the diameter of the
nebula have length/width ratios $\ga40$.

The density in the filaments can be estimated by measuring the
intensity enhancement ($\Delta I$) across a strand relative to the
average background emission of the nebula ($I$). Since the intensities
depend upon the emission measure along the line of sight, the electron
density in the filament ($n_{\rm f}$) is related to the average
density in the nebula $n_{\rm n}$ (assuming uniform conditions) by the
equation $n_{\rm f} = n_{\rm n} (\Delta I d_{\rm n} /I w_{\rm
f})^{1/2} $, where $w_{\rm f}$ is the filament width (ignoring
orientation effects) and $d_{\rm n}$ is the depth of the nebula along
the line of sight. Using the thickness of a filament given above, the
dimensions of the nebula from Table~1, and a filament/nebula intensity
contrast ($\Delta I/I$) of 10\%, which is representative of the
brighter filaments, we find $n_{\rm f} \sim 3n_{\rm n}$. Using the
value of $n_{\rm n}$ given in Table~1 and a temperature of $10^4$~K,
the gas pressure in the filaments is $1.0\times
10^{-7}$~dyne\,cm$^{-2}$.  IC~418 is not detected in molecular line
emission (Huggins et al. 1996), but it is seen in the \ion{H}{1} 21~cm
line (Gussie et al. 1995). The nebula is not resolved at 21~cm, so it is
uncertain if the filaments have  neutral atomic cores; Gussie et
al. (1995) interpret the \ion{H}{1} emission as arising from a neutral halo.

The global pattern of the filaments in IC~418 rules out shell
instabilities as a plausible formation scenario. The extreme
thread-like character of the filaments and their curved,
spaghetti-like geometry, which includes partial loops or helical
writhe, makes a strong case that they are magnetically dominated
structures. The filaments might be formed primarily from material in
the nebula shell, where it has been locally compressed to higher
density by the action of jets from the central star. However, the
narrow angle subtended at the center ($\sim 1\arcdeg$), which would
imply unusually high collimation for the jets, and the local geometry
probably rule this out.  The most likely formation scenario is one in
which the filaments were ejected over time as continuous or
intermittent strands, or multiple loops, from the central star.

\subsection{NGC~6537}

NGC~6537 is a bipolar/butterfly PN with an extremely hot central star
(Table~1). The nebula extends $\sim 120\arcsec$ from the center and is
formed by high velocity (300~\kms) outflows (Corradi \& Schwartz 1993)
which show complex structure.  The region of interest here is the
5\arcsec\ core of the nebula, which is shown in H$\alpha$ and
[\ion{N}{2}] $\lambda$6583 in Fig.~6.  The axis of the extended nebula
is roughly orthogonal to the axis of the core, but its structure
indicates that the outflows are directed at considerable angles from
the main axis, and emerge obliquely from the core region, as seen,
e.g., in ablated gas extending toward the top right in Fig.~6.

The core of NGC~6537 does not show such extreme filaments as the cases
discussed above, but is of interest as an example of a relatively
complete equatorial torus showing filamentary sub-structure.  The
torus is not axi-symmetric, but three main components can be
discerned: a kinked, main ring (roughly horizontal in the figure), and
two distorted, tilted, semi-circular filaments, linked or embedded in
the walls of the torus above and below the ring.  The structure seen
in H$\alpha$ and [\ion{N}{2}] is nearly the same; the main difference
being the absence of [\ion{N}{2}] emission from the center, where the
nitrogen is in higher ionization states because of the high stellar
temperature. Images in [\ion{O}{3}] $\lambda$5007 and [\ion{O}{1}]
$\lambda$6300 (not shown) are roughly similar.

The thickness of the filaments ranges from $\la {0}\farcs{2}$
(unresolved) to $\sim {0}\farcs{5}$ ($\la 7\times 10^{15}$~cm to
$1.8\times 10^{16}$~cm), and they subtend angles at the center of $\la
5\arcdeg$ to $\sim 12\arcdeg$, so they are generally larger structural
elements of the nebula than in the previous cases discussed. The
length/width ratio, assuming semi-circles with the radius of the
horizontal ring, ranges from 16 to $\ga 40$. The filaments dominate
the line emission from the core of NGC~6537 so the density in Table~1
is representative of the density in the bright ionized regions seen in
Fig.~6. The corresponding gas pressure is $2.8\times
10^{-8}$~dyne\,cm$^{-2}$. The central regions of NGC~6537 show CO
emission (Huggins et al. 2005) which probably implies that the
filaments have molecular cores below the photo-ionized surfaces.

The expansion velocity of the torus is 18~\kms\ (Table 1) and its
expansion time scale (1,600~yr) is comparable to that of the much more
extended nebula formed by the high velocity outflows. The torus
represents the last major mass ejection by the central star.  The
ejection was sudden, there is no radially extended equatorial disk,
and the mass is substantial, $\sim 0.02$~\Msun\ in ionized gas alone
(estimated from the parameters in Table~1). The low velocity of the
torus and the continuity of the filaments over large angles in
longitude suggests that they were formed during the ejection
process. Their distortion is probably a result of interaction with the
high velocity outflows since the upper filament is evidently being
ablated. The arguments that the filaments are held together with a
magnetic field follow those for NGC~3132.  The conclusion is less
compelling than in the previous cases where the filamentary structures
have such extreme geometries, but is still the most reasonable
explanation of their structure.

\section{Discussion}

\subsection{ Field Strengths}

Our interpretation of the geometry of the filaments described above is
that they provide morphological evidence for the influence of magnetic
fields in the extended structure of PNe.  If this is correct, the
properties of the filaments can be used as a guide to the likely field
strengths.

The most important consideration for the integrity of a filament in
the expanding PN environment is its connectivity. Using a very simple
picture ignoring the effects of vorticity, the breaking time of a
non-magnetic, tangential filament is given by $t_{\rm b} = w_{\rm
f}/\Delta V$ where $\Delta V$ is the velocity of any radial shear.
For $\Delta V = V_{exp}$ ($\sim 18$~\kms), $t_{\rm b}$ is 50~yr, much
shorter than the expansion time of the nebulae ($\sim 3,000$~yr). Even
if $\Delta V$ is as low as 1~\kms, which may be typical of the
turbulent velocity in the cool precursor envelope (e.g., Huggins \&
Healy 1986), $t_{\rm b}$ is still less than the expansion time
scale. Thus, unless the filament is connected along its length, it is
likely to break.

A second consideration for the integrity of a filament is its lateral
dispersion. However, this is a less crucial issue. From the observations we
cannot determine the extent to which a filament is in close pressure
equilibrium with its surroundings, so for illustration we consider
free expansion, for which the dispersion time scale is $t_{\rm d} =
w_{\rm f}/2c_{\rm s}$ where $c_{s}$ is the sound speed.  Although
$t_{\rm d}$ is short ($\sim 50$~yr) compared with the expansion time
scales of the nebulae if the gas is ionized ($c_{s}\sim 10$~\kms), it
is probable that the filaments of NGC~3132 and NGC~6537, and possibly
IC~418, have neutral, low temperature cores with low sound speeds. In
this case the ionized gas seen in the images forms a thin skin or a
photo-evaporation flow from the surface of the filament. Even if this
gas is not confined and disperses on short time scales, a filament can
maintain its thin lateral dimension as long as the ionization front is
weak, and the cool, neutral core is maintained.

Two standard parameters describe the importance of a magnetic field in
the PNe and are relevant to the integrity of a filament: $\beta$, the
ratio of the thermal gas pressure to the magnetic pressure, defined in
section~1, and $\sigma$, the ratio of the magnetic energy density to
the kinetic energy density, i.e., $\sigma = B^2/(4 \pi \rho V_{\rm
exp}^{2})$ where $\rho$ is the gas density.  Using the gas densities
and gas pressures in the filaments determined in the previous sections
and the values of $V_{\rm exp}$ from Table~1, we calculate the
strength of the fields ($B_{\rm P}$) for which ${\beta} = 1$, and the
strength of the fields ($B_{\rm E}$) for which $\sigma =1$. The
results are given in Table~3, together with representative distances
of the filaments from the central stars.  It can be seen from the
table that these values of the fields are not very different from one
another ($\sim 1$~mG).

A field with $\sigma \sim 1$ would dominate the overall dynamics of a
filament and would render it extremely stiff against perturbations.  A
field with $\beta \sim 1$ would also play a role in the local behavior
of a filament, and with the appropriate helical geometry would help or
dominate the confinement (e.g., Fiege \& Pudritz 2000). However,
fields estimated on the basis of equi-partition are probably upper
limits because of the high magnetic tension (see also Soker 2005). The
magnetic tension force is given by $T_{\rm M} = B^2/4 \pi R_{\rm c}$
per unit volume, where $R_{\rm c}$ is the radius of curvature of the
field.  For a filament at radius $r$ expanding with velocity $V_{\rm
exp}$, the tension force can be expressed as $T_{\rm M} = \sigma \rho
V_{\rm exp}^{2}/r $, and for $\sigma = 1$, it can be seen to be
sufficiently large to arrest the motion of the filament on a time
scale equal to the expansion time scale ($r/V_{\rm exp}$), unless
there is also a strong, outwardly directed gradient in the field.  The
observed filaments do not currently appear to have a separate
dynamical evolution from other parts of the nebulae. 

A more realistic scenario is one in which the magnetic field plays a
dynamical role closer to the central star where the fields might be
relatively stronger.  For example, $\sigma \la 1$ at some
representative radius $r_{\rm a}$ where the filaments are formed
and/or the gas is accelerated to its terminal velocity. We do not know
how the field has varied with distance from the star, but if we assume
that the gas density at $r_{\rm a}$ is related to the observed density
by $ \rho_{\rm a} = \rho (r/r_{\rm a})^2 $ and the velocity is $\sim
V_{\rm exp}$, we can estimate the equi-partition field at $r_a$.  The
results for a representative radius of $r_{\rm a} = 10^{15}$~cm are
given in Table~4, labeled $B_{\rm E}(10^{15})$.

An alternative scenario is one in which the magnetic field plays a
subordinate role in the overall dynamics, but the magnetic tension
plays a cohesive role on small length scales (see also Soker
2002). From the expression given above ($T_{\rm M} = \sigma \rho
V_{\rm exp}^{2}/r $), it can be seen that even with $\sigma \ll 1$ the
tension that results from small scale ($\ll r$) perturbations can be
large.  For example, if a filament is subject to disruptive forces
that would produce variations in the terminal velocity of $\sim V_{\rm
exp}/10$ on length scales of $\sim r/10$ in an expansion time, the
disrupting forces can be effectively balanced by the tension if the
field is $\ga B_{\rm E}(r)/10$. Thus fields significantly smaller than
equi-partition fields can still provide strong, local connectivity
along the length of a filament.

\subsection{Comparison with Observed Fields}

The magnetic fields estimated on the basis of equi-partition in the
previous section are plotted against radius in Figure~7.  For
comparison we show the fields measured in the maser spots of
circumstellar envelopes, as presented by Vlemmings et al.\ (2002). The
boxes show the fields from SiO and OH masers in Mira variables (dashed
lines) and supergiants (solid lines), and the dashed and solid lines
indicate corresponding $B \sim r^{-2}$ fits to the radial variation of
the field.  The typical precursor of a PN at the tip of the AGB
probably lies somewhere near the supergiant data. The OH masers of the
two proto-PNe discussed by Bains et al. (2004) have fields of $\sim
3$~mG at radii of $6\times10^{15}$~cm and $3\times10^{16}$~cm, and lie
inside the supergiant OH maser box. Recent observations of H$_2$O
masers by Vlemmings et al.\ (2005) suggest a dipole ($r^{-3}$) field
rather than an $r^{-2}$ field in the supergiant VX Sgr, but this does
not change the overall picture presented here.

Could the PN filaments have developed from the fields seen in the
precursor envelopes? It can be seen from Fig.~7 that the current
equi-partition fields in the filaments are comparable to those in the OH
masers in the envelopes, and they lie farther from the central star by
about an order of magnitude. They therefore lie above the solid line
in the figure by 1--2 orders of magnitude.  Although we regard the
equi-partition fields as likely overestimates, fields of this
magnitude can probably be produced at their current distances from the
star by the compression of the neutral gas in the AGB envelopes when
overtaken by the ionization front of the nebula, with a corresponding
increase in the field (e.g., Soker 2002).  More importantly, when we
consider the field estimates at the acceleration region closer to the
star, the equi-partition fields of the PN filaments are comparable to
the observed maser fields, and subordinate but cohesive fields (as
discussed above) are even lower. We therefore conclude that the fields
needed to play a role in the development of PN filaments could have
originated in the precursor circumstellar gas.

For completeness, we note that a PN central star with an average
surface field of 1.5~kG and a radius 0.3~$R_{\odot}$, comparable to
the cases recently reported by Jordan et al.\ (2005), would likely
produce weak fields in the extended nebula.  For example, a variation
of $r^{-2}$ from the surface would fall about three orders of
magnitude below the dashed line in Fig.~7.  This could be increased by
a less steep ($r^{-1}$) radial dependence and/or shock compression,
but the relation of the observed stellar field to that of the AGB
stars is not yet clear.  The geometry suggested by the filaments may
be important in developing this picture.

\subsection{Geometry}

Our suggestion that the filaments may be identified with the
underlying magnetic field structure in the nebulae is of interest
because there are currently no other guides to the detailed geometry
of the fields in PNe.  The PNe considered here have different overall
morphologies, but they share a common characteristic of localized
structures with long-range longitudinal coherence, i.e., they are
stringy. This is rather different from the relatively smooth fields
that are typically used in theoretical models of magnetic fields in
PNe and may indicate that high resolution simulations are needed to
produce realistic models.  Of course the PNe were selected on the
basis of their filaments, so they may be isolated, peculiar
cases. However, they are all relatively nearby objects, and their
other properties are quite typical of PNe, so they may be
representative of a wide distribution of the PN population.

In all three PNe the filaments are roughly perpendicular to the radius
vector. This is consistent with the dominance of tangential magnetic
fields at large distances from the central stars as expected on
general theoretical considerations (e.g., Pascoli et al.\ 1992;
Chevalier \& Luo 1994).  The filaments in IC~418 and NGC~3132 are very
thin with large length-to-width ratios, and this may be also be the
case in NGC~6537 but it is more distant so that any small scale
structure is not resolved in the images. The filaments in NGC~6537 and
NGC~3132 are parts of rings or tori, which are made up of more than
one filament. Rings and tori are very common in PNe, and they may have
similar substructure when observed at high spatial resolution.  The
case of IC~418 is different in that it involves time dependent changes
in the direction of ejection, but the underlying mechanisms may be similar.

There are no detailed theoretical predictions of the local geometry of
the magnetic field in PNe.  However, based on analogy with the case of
the solar field and magnetic clouds (e.g., Russell, Priest, \& Lee
1990) and models for filamentary clouds in the interstellar medium
(e.g., Fiege \& Pudritz 2000), we expect that the filaments are in
the form of magnetic flux ropes, with longitudinal fields in the core,
surrounded by azimuthal components farther from the axis. It may be
possible to investigate this directly with additional observations.

The geometry of the filaments and the likely field strengths discussed
in previous sections lead naturally to the possibility that the OH
maser spots in circumstellar envelopes may be regions of enhanced gas
density and magnetic field associated with proto-filaments. The recent
suggestion from a different perspective by Soker \& Kastner (2003)
that the maser spots in AGB envelopes represent local
enhancements of the magnetic field, is consistent with this idea.
However, there is a sharp transition from roughly spherical mass loss
on the AGB to toroidal structures in proto-PNe: if magnetic fields
play an important role in this transition, the filaments that we
describe here, especially the toroidal ones, could also be the
products of this enhanced magnetic activity.

\section{Conclusions}

The HST observations reported in this paper provide a means of
exploring the potential connection between filaments and magnetic
fields in PNe, a possibility mentioned two decades ago by Pascoli
(1985).  The high resolution images provide striking examples of
extreme filamentary structures.  The thread-like character and
geometry of the filaments lead us to propose that they may be
signatures of the underlying magnetic field.  We have derived the
physical properties of the gas in the filaments, and find that fields
which would be strong enough to provide coherence, are probably
consistent with those measured in the precursor circumstellar
envelopes.  The filaments suggest that the magnetic fields in PNe may
be localized and thread-like.

\acknowledgements
We thank Dr.\ Adam Frank and an anonymous referee for helpful
comments.  This work has been supported in part by NSF grant
AST~03-07277 (to P.J.H.).


\clearpage

\begin{deluxetable}{llccrrcl}
\tablecaption{PN Properties}
\tablewidth{0pt}
\tablehead{
\colhead{PN} & \colhead{Name} & \colhead{$d$} & \colhead{$\theta_{\rm
D}$} & \colhead{$n_{\rm e}$} &
\colhead{$T_*$} & \colhead{$V_{\rm exp}$} & \colhead{$t_{\rm exp}$} \\
 & & \colhead{(kpc)} & \colhead{($\arcsec$)} & \colhead{(cm$^{-3}$)} &
\colhead{(K)} & \colhead{(km\,s$^{-1}$)} & \colhead{(yr)} 
}
\startdata
010.1$+$00.7 & NGC 6537 & 2.4 & 10 & 10,000 & $\ga$500,000 & 18 & 3,200 \\
215.2$-$24.2 & IC 418   & 1.0 & 12 & 12,000 &   33,000     & 12 & 2,400 \\
272.1$+$12.3 & NGC 3132 & 0.6 & 56 & 700    &    99,000    & 21 & 3,800 \\

\enddata




\end{deluxetable}
\clearpage

\begin{deluxetable}{lccc}
\tablecaption{Image Information}
\tablewidth{0pt}
\tablehead{
\colhead{PN} & \colhead{Dataset} & \colhead{Filter}  &
\colhead{Exp. Time}  \\
 &  &  & \colhead{(s)} 
}
\startdata
NGC 3132 &  U5HC0101B &  F502N  &  800  \\
         &  U5HC0103B &  F658N  &  700  \\
IC 418   &  U35T0905B &  F656N  &  880  \\
         &  U66B1006B &  F658N  &  700  \\
NGC 6537 &  U42I0402B &  F656N  &  1240  \\
         &  U42I0407B &  F658N  &  1200  \\
\enddata
\end{deluxetable}

\clearpage

\begin{deluxetable}{lcccc}
\tablecaption{Equi-partition Magnetic Fields}
\tablewidth{0pt}
\tablehead{
\colhead{PN} & \colhead{$r$} & \colhead{$B_{\rm P}(r)$}  &
\colhead{$B_{\rm E}(r)$} & \colhead{$B_{\rm E}(10^{15})$} \\
 & \colhead{($10^{16}$ cm)} & \colhead{(mG)} & \colhead{(mG)} &  \colhead{(mG)}
}
\startdata
NGC 6537 &  9 &  0.84  &  0.82 & 74 \\
IC 418   &  9 &  1.60  &  1.00 & 94 \\
NGC 3132 & 18 &  0.21  &  3.47 & 624 \\

\enddata
\end{deluxetable}

\clearpage




\figcaption[]{Overview image of NGC~3132, adapted from the Hubble
Heritage image. The field is $87\arcsec \times 61 \arcsec$. North is
$109\arcdeg$ counter-clockwise from vertical. 
 }

\vspace{1.0cm}
\figcaption[]{Filaments in NGC~3132. \emph{Left}: Images in
[\ion{O}{3}] $\lambda$5007 (top) and [\ion{N}{2}] $\lambda$6583
(bottom) with logarithmic intensity scales. \emph{Right}:
Unsharp mask versions of these images (see text for details). The
fields are $12\arcsec \times 19\arcsec$.
}

\vspace{1.0cm}
\figcaption[]{Intensity cuts in [\ion{O}{3}] $\lambda$5007 through the
filaments of NGC~3132. The strips are horizontal cuts through the
image in Fig.~2. They are labeled with the vertical distance in arc
seconds from the knot feature, and are offset along the intensity axis
for clarity.
}

\vspace{1.0cm}
\figcaption[]{Overview image of IC~418 in H$\alpha$. The field is 
$18\arcsec \times 18\arcsec$. North is $117\arcdeg$ clockwise from
vertical.
}

\vspace{1.0cm}
\figcaption[]{Filaments in IC~418.  \emph{Top left}: H$\alpha$, field
$18\arcsec \times 18\arcsec$. \emph{Top right}: H$\alpha$, field
$4\farcs{7} \times 4\farcs{7}$. \emph{Bottom right}: [\ion{N}{2}]
$\lambda$6583, field $4\farcs{7} \times 4\farcs{7}$. \emph{Bottom
left}: H$\alpha$, field $2\farcs{9} \times 2\farcs{9}$. All panels are
unsharp mask images, smoothed with a FWHM of 2 pixels.
}

\vspace{1.0cm}
\figcaption[]{Images of the torus of NGC~6537 in H$\alpha$ (top) and
[\ion{N}{2}] $\lambda$6583 (bottom). The greyscale is inverted. 
The field is $9\arcsec \times
8\arcsec$, north is $45\arcdeg$ counter-clockwise from vertical.
}

\vspace{1.0cm}
\figcaption[]{Magnetic field strength versus radius for evolved stars
and PNe. The equi-partition fields of the PNe from Table~3 are denoted
by crosses ($B_{\rm P}$), filled triangles ($B_{\rm E}$ ) and open
triangles ($B_{\rm E}$ at 10$^{15}$~cm).  The boxes denote the ranges
for SiO and OH masers for Miras (dashed lined) and supergiants (solid
line), and the solid and dashed lines are $r^{-2}$ fits through these
from Vlemmings et al.\ (2002). 
}

\clearpage
\begin{figure}
\begin{center}
\resizebox{12.0cm}{!}{\includegraphics{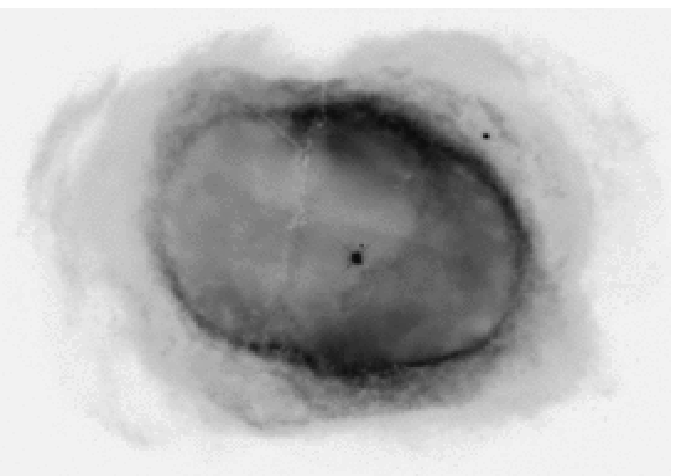}}
\end{center}
\end{figure}
\clearpage

\clearpage
\begin{figure}
\begin{center}
\resizebox{!}{11.0cm}{\includegraphics[angle=0]{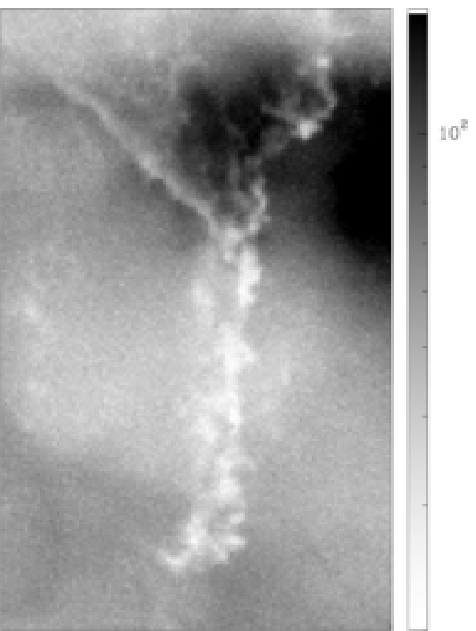}}
\hspace{0.6cm}
\resizebox*{!}{11.0cm}{\includegraphics[angle=0]{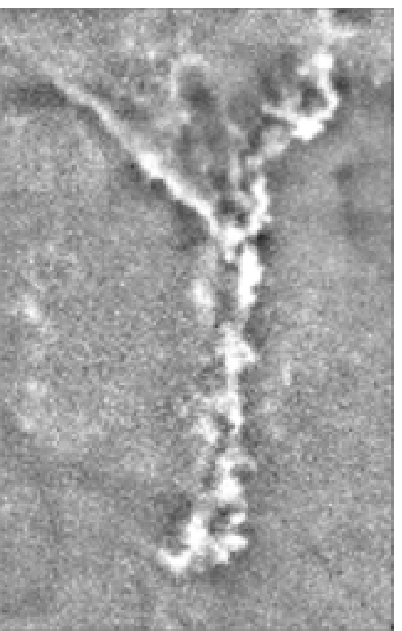}}

\vspace{1.2cm}
\resizebox{!}{11.0cm}{\includegraphics[angle=0]{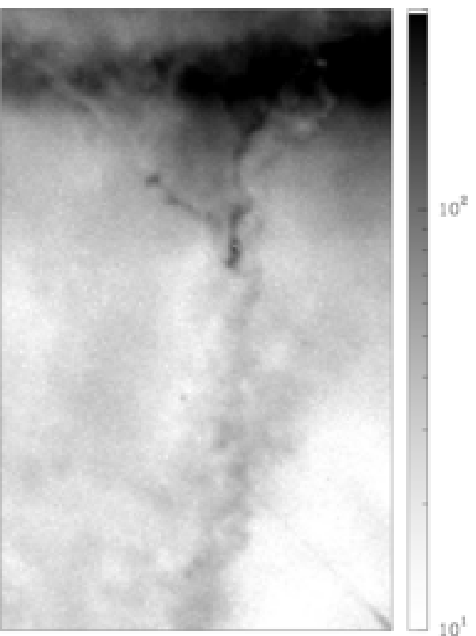}}
\hspace{0.6cm}
\resizebox{!}{11.0cm}{\includegraphics[angle=0]{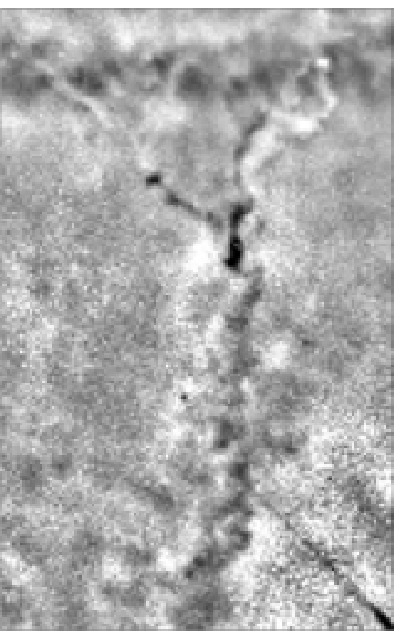}}
\end{center}
\end{figure}

\clearpage
\begin{figure}
\begin{center}
\resizebox{10.0cm}{!}{\includegraphics[angle=0]{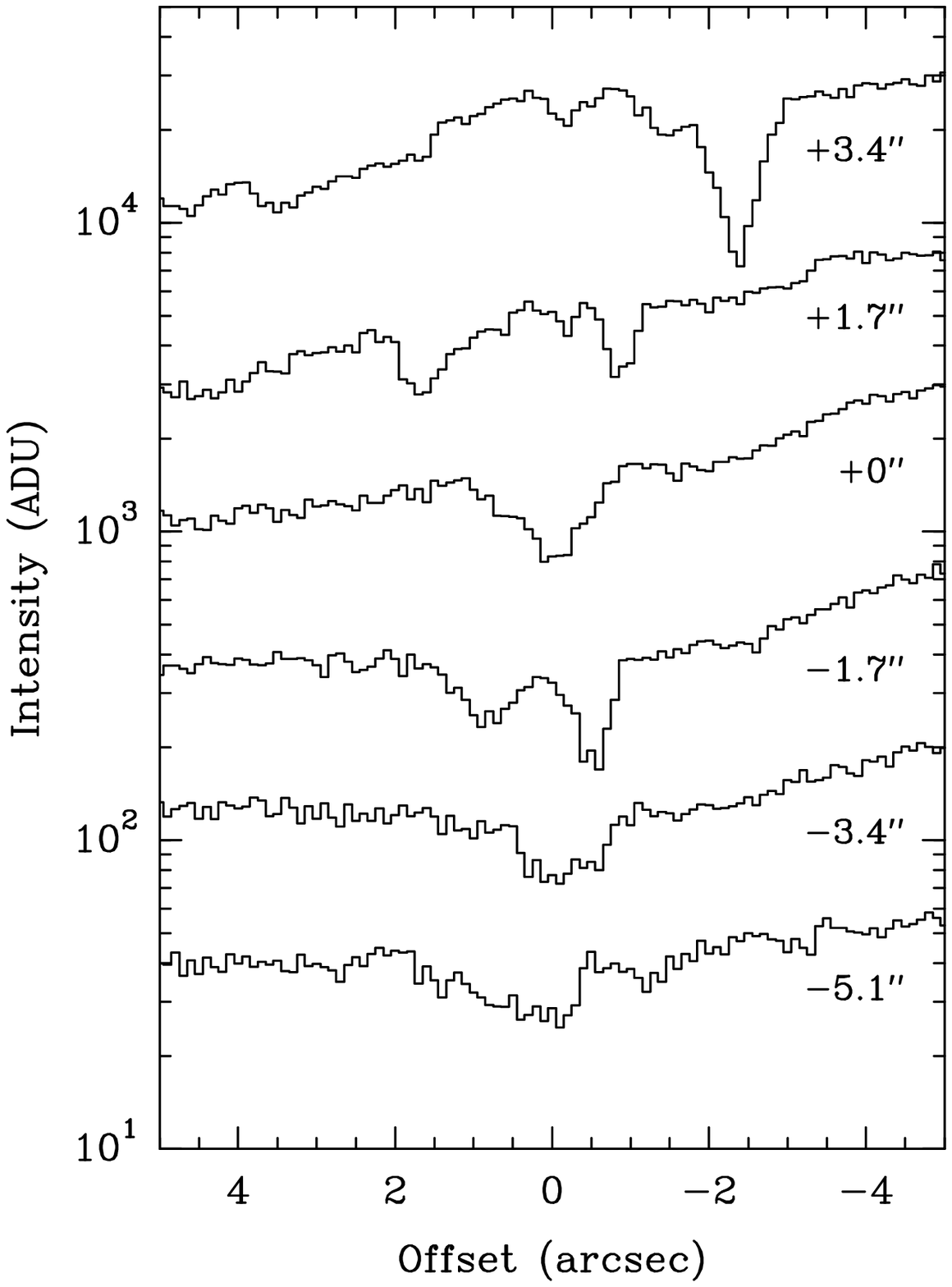}}
\end{center}
\end{figure}

\clearpage
\begin{figure}
\begin{center}
\resizebox{13.0cm}{!}{\includegraphics{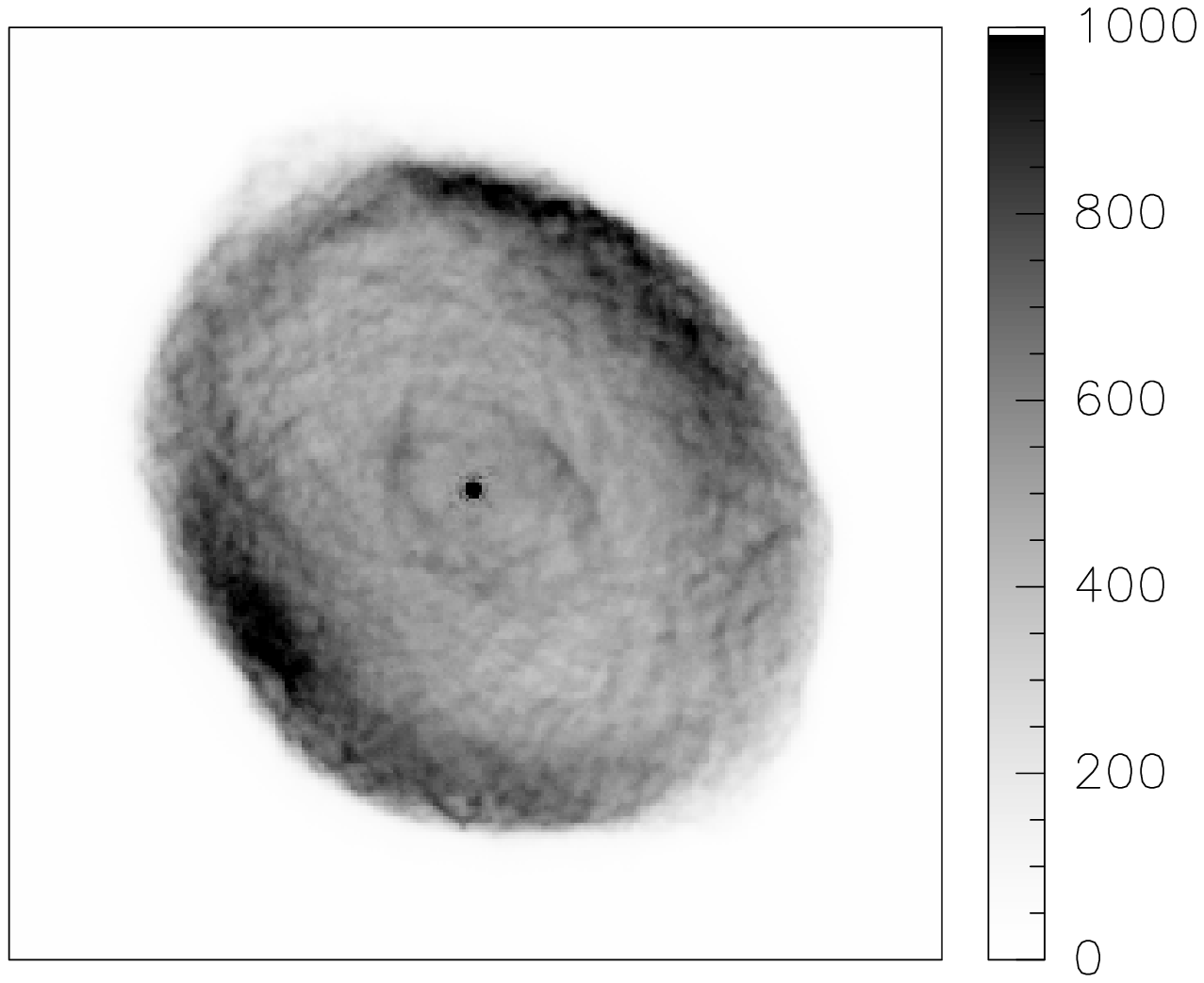}}

\end{center}
\end{figure}

\clearpage
\begin{figure}
\begin{center}
\resizebox{7.5cm}{!}{\includegraphics{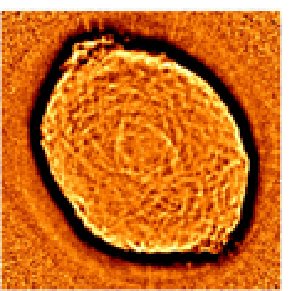}}
\hspace{0.20cm}
\resizebox{7.5cm}{!}{\includegraphics{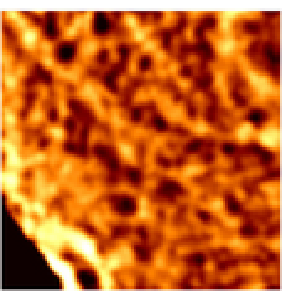}}

\vspace{0.5cm}
\resizebox{7.5cm}{!}{\includegraphics{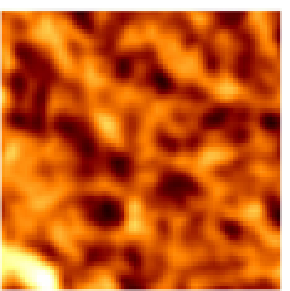}}
\hspace{0.20cm}
\resizebox{7.5cm}{!}{\includegraphics{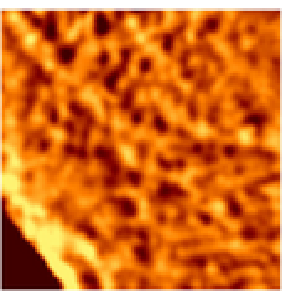}}
\end{center}
\end{figure}

\clearpage
\begin{figure}
\begin{center}
\resizebox{12.0cm}{!}{\includegraphics[angle=0]{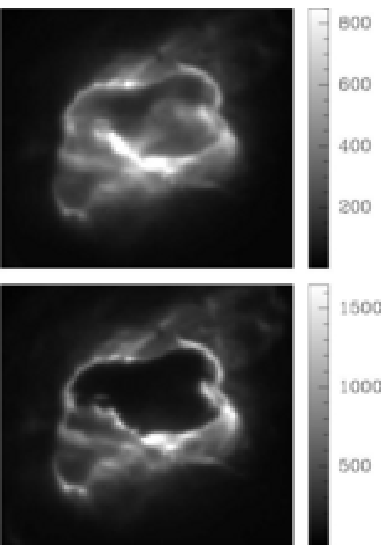}}
\end{center}
\end{figure}

\clearpage
\begin{figure}
\begin{center}
\resizebox{14.0cm}{!}{\includegraphics[angle=0]{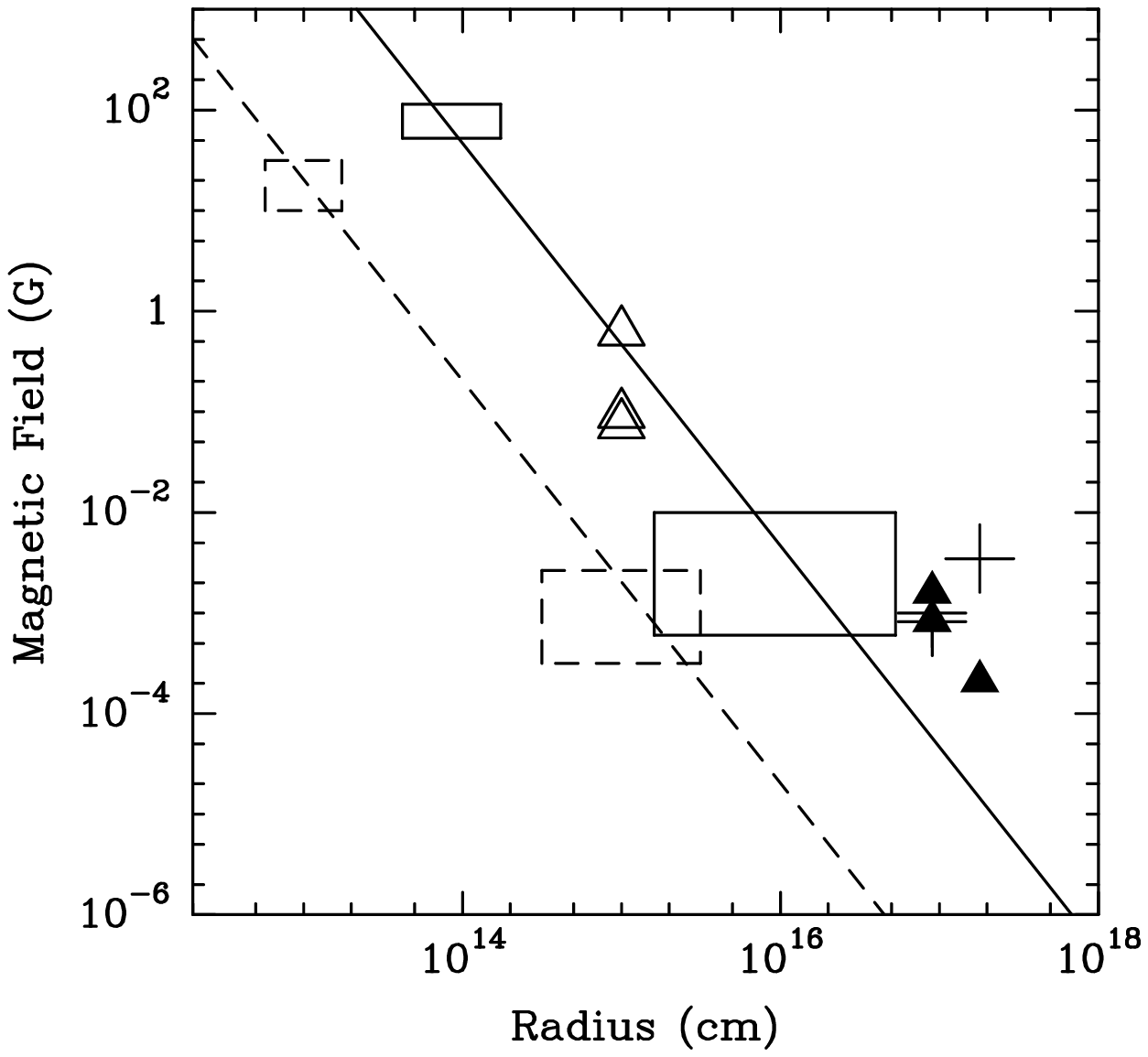}}
\end{center}
\end{figure}

\end{document}